\documentclass[twocolumn,aps,pra,linenumber,showpacs,floatfix,superscriptaddress]{revtex4-1}
\usepackage{amsfonts}
\usepackage{amsmath}
\usepackage{amssymb}
\usepackage{dcolumn}
\usepackage{graphicx}%
\usepackage{bm}
\usepackage{color}

\begin{document}

\newcommand{\HIM}{
Helmholtz Institute Mainz, D-55128 Mainz, Germany}

\newcommand{\NZIAS}{
Centre for Theoretical Chemistry and Physics,
The New Zealand Institute for Advanced Study,
Massey University Auckland, Private Bag 102904, 0745 Auckland, New Zealand}

\newcommand{\UNSW}{
School of Physics, University of New South Wales, Sydney 2052, Australia}

\newcommand{\MBU}{
Department of Chemistry, Faculty of Natural Sciences, Matej Bel University, 
Tajovsk\'{e}ho 40, SK-974 00 Bansk\'{a} Bystrica, Slovakia}

\newcommand{\PUM}{Fachbereich Chemie, Philipps-Universit\"{a}t Marburg, 
Hans-Meerwein-Str., D-35032 Marburg, Germany}

\newcommand{\FIAS}{Frankfurt Institute for Advanced Studies, Ruth-Mounfang-Str.1, 60438, Frankfurt am Main, Germany}


\title{Relativistic study of the nuclear anapole moment effects in diatomic molecules} 


\author{A. Borschevsky}
\affiliation{\HIM}
\affiliation{\NZIAS}

\author{M. Ilia\v{s}}
\affiliation{\MBU}

\author{V. A. Dzuba}
\affiliation{\UNSW}


\author{V. V. Flambaum}
\affiliation{\UNSW}
\affiliation{\NZIAS}
\affiliation{\FIAS}

\author{P. Schwerdtfeger}
\affiliation{\NZIAS}
\affiliation{\PUM}

\date{\today}


\begin{abstract}


Nuclear-spin-dependent (NSD) parity violating effects are studied for a number of diatomic
molecules using relativistic Hartree-Fock and density functional theory and accounting for core polarization
effects. Heavy diatomic molecules are good candidates for the successful measurement of the
nuclear anapole moment, which is the dominant NSD parity violation term in heavy elements.
Improved results for the molecules studied in our previous publication [Borschevsky \textit{et al.}, Phys.
Rev. A \textbf{85}, 052509 (2012)] are presented along with the calculations for a number of new promising candidates for the
nuclear anapole measurements.

\end{abstract}

\maketitle

\section{Introduction}
 
The parity ($P$)--odd and time ($T$)--even nuclear anapole moment originates from the second-order term in the multipole expansion of the 
magnetic vector-potential together with the $P$-- and $T$-- violating  magnetic quadrupole moment \cite{SusFlaKri84}. 
It provides the dominant contribution to the nuclear-spin-dependent (NSD) parity violating (PV) effect in atoms and molecules \cite{FlaKri80}.
The corresponding term in the Hamiltonian arising from this NSD PV electron-nucleus interaction for a single electron 
is (atomic units are used throughout)
\begin{eqnarray}
H_A=\kappa_{\rm NSD} \frac{G_F}{\sqrt{2}}\frac{\bm{\alpha}\cdot \mathbf{I}}{I}\rho(\mathbf{r}),
\label{Ha}
\end{eqnarray}
where $\kappa_{\rm NSD}$ is the dimensionless strength constant, $G_{F}=2.22249\times10^{-14}$ a.u. the Fermi coupling
constant, $\bm{\alpha}$ is the vector comprised of the conventional Dirac matrices, $\mathbf{I}$ is the
nuclear spin, $\mathbf{r}$ is the displacement of the valence electron from
the nucleus, and $\rho(\mathbf{r})$ is the (normalized) nuclear density.
The nuclear anapole moment contribution to the NSD interaction requires nuclear spin $I\neq 0$ and 
in a simple  valence  model  has the following value  \cite{FlaKriSus84}
\begin{eqnarray}
\kappa_{\rm A}= 1.15\times 10^{-3} \left(\frac{\mathcal{K}}{I+1} \right) A^{2/3} \mu_m g_m,
\label{eqanapole}
\end{eqnarray}
Here, $A$ is the number of nucleons, $\mathcal{K}=(-1)^{I+\frac{1}{2}-l}(I+1/2)$, $l$ is the orbital
angular momentum of the external unpaired nucleon, i.e. $m=n,p$; $\mu_p= +2.8$, $\mu_n= -1.9$. Theoretical estimates give the strength constant for
nucleon-nucleus weak potential $g_p \approx +4.5$ for a proton and
$|g_n|\sim 1$ for a neutron \cite{FlaMur97}.
Since the nuclear anapole moment contribution to the NSD interaction scales with the number of nucleons, $\kappa_{A}\sim A^{2/3}$, it becomes the dominant contribution in spin-dependent atomic parity violation effects for sufficiently large nuclear charge $Z$ \cite{FlaKri80,FlaKriSus84}. Two other contributions to the NSD interaction arise from the electroweak neutral coupling between the electron vector and 
the nucleon axial-vector currents ($\mathbf{V}_e\mathbf{A}_N$) \cite{NovSusFla77} and from the nuclear-spin-independent weak interaction combined with the hyperfine interaction \cite{FlaKri85}. 

The nuclear anapole moment was experimentally measured  
in the $^{133}$Cs atom in 1997 \cite{WooBenCho97}, following a proposal by Flambaum and
Khriplovich \cite{FlaKri80}. Further anapole measurements are however required to provide accurate values for the $g_p $ and $g_n$ constants, thus obtaining important information about hadronic weak coupling.

In Refs. \cite{Lab78,SusFla78,FlaKri85_2} it was shown that the
nuclear spin-dependent parity violation effects are enhanced by a
factor of $10^5$ in diatomic molecules with  $^2\Sigma_{1/2}$ and
$^2\Pi_{1/2}$ electronic states due to the mixing of close rotational
states of opposite parity ($\Omega$-doublet for
$^2\Pi_{1/2}$). These systems are thus advantageous for measurement of the NSD parity violation effects, and should provide data on anapole moments for many heavy nuclei. Two experimental proposals for using diatomic molecules to measure NSD parity violation effects were published in recent years \cite{DemCahMur08,IsaHoeBer10}, and the corresponding experiments have already began. An additional experiment of precision spectroscopy of the cooled SrF molecule is conducted at the University of Groningen \cite{Hoe13}. Diatomic positive molecular ions also experience the enhancement of the NSD PV effects, and have an additional experimental advantage of being easy to trap and study at low temperatures \cite{Odom11}. An experiment to cool diatomic molecular ions in the $^2\Sigma_{1/2}$ ground state for measurement of PV effects is currently attempted at the Northwestern University \cite{Odo13}.
 
A number of theoretical investigations of the NSD parity violation in neutral diatomic molecules have been published in recent years, using both semiempirical \cite{KozFomDmi87,DmiKhaKoz92,KozLab95,DemCahMur08} and \textit{ab-initio}   \cite{TitMosEzh96,KozTitMos97,NayDas09,BakPetTit10,IsaHoeBer10,IsaBer12} methods. These calculations focused on obtaining the NSD P-odd interaction constant $W_A$ that is necessary to relate the experimental measurements
to the $\kappa_A$ constant arising from the NSD PV interaction within the nucleus \cite{HaxWie01}.
In two recent publications we presented Dirac Hartree-Fock (DHF) and relativistic density-functional (DFT) calculations 
of the $W_{A}$ factor of a variety of neutral diatomic molecules \cite{BorIliDzu12} and diatomic positively charged molecular ions \cite{BorIliDzu12_2}. Here we present improved calculations of the  $W_{A}$ factor for the molecules discussed in Ref \cite{BorIliDzu12}, obtained using a more accurate treatment of the finite nuclear size
and the basis sets. In addition, we present the  $W_{A}$ parameters for a number of systems which have not been previously studied in the context of NSD PV, namely group 3 oxides, group 4 nitrides, and group 14 fluorides.

\section{Computational Details}
For $^2\Sigma_{1/2}$ and $^2\Pi_{1/2}$ electronic states considered, the interaction (\ref{Ha}) can be replaced by the effective operator, 
which appears in the spin-rotational Hamiltonian \cite{FlaKri85_2,DemCahMur08},
\begin{eqnarray}
H_A^\mathrm{eff}=\kappa_{\rm NSD} W_A\frac{(\mathbf{n}\times\mathbf{S}^\prime)\cdot \mathbf{I}}{I},
\label{eq:Heff}
\end{eqnarray}
where $\mathbf{S}^\prime$ is the effective spin and
$\mathbf{n}$ is the unit vector directed along the molecular axis  
from the heavier to the lighter nucleus. The electronic factor $W_A$ is found from evaluating the matrix
elements of the $\bm{\alpha}\rho(\mathbf{r})$ operator in the
molecular spinor basis \cite{Visscher1997181}. 
The $^2\Sigma_{1/2}$ and the $^2\Pi_{1/2}$ open-shell electronic states are
two-fold degenerate, corresponding to the two possible projections of
electronic angular momentum along $\mathbf{n}$,
i.e.~$|\Omega\rangle=|\pm\frac{1}{2}\rangle$.  
When operating within this degenerate space, the operator
$\frac{G_F}{\sqrt{2}}\bm{\alpha}\rho(\mathbf{r})$ is equivalent to
$W_A(\mathbf{n}\times\mathbf{S}^\prime$) (Eq.~(\ref{eq:Heff})).  
Time-reversal symmetry ensures that only the matrix elements that are
off-diagonal in $\Omega$ are non-vanishing. This symmetry rule is
encapsulated within the effective operator $H_A^\mathrm{eff}$ by the
angular factor $(\mathbf{n}\times\mathbf{S}^\prime)$. Here the
effective spin $\mathbf{S}^\prime$ generates rotations in the
degenerate subspace analogously to the usual spin operator $\mathbf{S}$ in
a spin-1/2 system. 

The calculations
were carried out within the open-shell single determinant
average-of-configuration Dirac-Hartree-Fock approach
\cite{Thyssen_thesis} and within the relativistic density functional theory \cite{SauHel02}, employing quaternion  
symmetry \cite{Saue:1997, Saue:1999}. A finite nucleus, modeled by the Gaussian 
charge distribution was used \cite{VisDya97}. In the DFT calculations we used the Coulomb-attenuated B3LYP functional 
(CAMB3LYP*), the parameters
of which were adjusted by Thierfelder \textit{et al.} \cite{ThiRauSch10} to reproduce the PV energy shifts obtained
using coupled cluster calculations (the adjusted parameters are $\alpha= 0.20$,
$\beta=0.12$, and $\mu=0.90$). All the calculations were performed using the DIRAC12 program package \cite{DIRAC12}. 

For the lighter elements (N to Si), the aug-cc-pVTZ basis sets were
used \cite{KenDunHar92,WooDun93}, all in their uncontracted form,
as it is important to have maximum flexibility of the wavefunction in the
region close to the nucleus. For the rest of the atoms, we employed the
Faegri's dual family basis sets \cite{Fae01}. As a good description of the electronic
wave function in the nuclear region is essential for obtaining reliable
results for parity violating properties \cite{LaeSch99}, we
augmented the basis sets with high exponent $s$ and $p$ functions, which
brings about an increase of around $10\%$ in the calculated values of $W_{A}$.
The basis sets were increased, both in the core and in the valence regions, to
convergence with respect to the calculated $W_{A}$ constants. The final basis sets can be found in Table \ref{tab:I}.

Where available, experimental bond distances $R_{e}$ were used. 
For molecules where $R_{e}$ is not known experimentally we optimized the bond
distance using the relativistic coupled cluster
approach with single, double, and perturbative triple excitations,
CCSD(T) \cite{Visscher:1996}. To reduce the computational effort,
we employed an infinite order two-component relativistic
Hamiltonian obtained after the Barysz--Sadlej--Snijders (BSS) transformation of
the Dirac Hamiltonian in a finite basis set \cite{IliJenKel05,Ilias:2007}. 
Our calculated $R_{e}$ are typically within 0.01 \AA \ of the experimental values, where available.

\begin{table}
  \caption{Basis sets employed in the calculations of the $W_{A}$ constants. All
elements with $Z>15$ are described by the Faegri basis sets \cite{Fae01}
augmented by high exponent, diffuse, and high angular momentum functions.}
  \label{tab:I}
  \centering
\begin{tabular}
[c]{lrr}\toprule
Atom\ \ \  &\ \ \ \  $Z$ &\hspace{2.5cm} Basis Set\\\colrule
N & 7 & aug-cc-PVTZ\\
O & 8 & aug-cc-PVTZ\\
F & 9 & aug-cc-PVTZ\\
Mg & 12 & aug-cc-PVTZ\footnote{augmented by 2 high exponent $s$ and 4 high exponent $p$ functions.}\\
Si & 14 & aug-cc-PVTZ\footnotemark[1]\\
Ca & 20 & 20\textit{s}18\textit{p}9\textit{d}6\textit{f}1\textit{g}\\
Sc & 21 & 19\textit{s}17\textit{p}10\textit{d}7\textit{f}2\textit{g}\\
Ti & 22 & 21\textit{s}16\textit{p}10\textit{d}7\textit{f}2\textit{g}\\
Zn & 30 & 21\textit{s}19\textit{p}10\textit{d}7\textit{f}2\textit{g}\\
Ge & 32 &  20\textit{s}20\textit{p}11\textit{d}8\textit{f}2\textit{g}\\
Br & 35 & 21\textit{s}20\textit{p}10\textit{d}10\textit{f}1\textit{g}\\
Sr & 38 & 21\textit{s}20\textit{p}12\textit{d}9\textit{f}2\textit{g}\\
Y & 39 & 21\textit{s}20\textit{p}12\textit{d}9\textit{f}2\textit{g}\\
Zr & 40 & 21\textit{s}20\textit{p}12\textit{d}9\textit{f}2\textit{g}\\
Cd & 48 & 22\textit{s}20\textit{p}12\textit{d}9\textit{f}2\textit{g}\\
Sn & 50 &  21\textit{s}21\textit{p}12\textit{d}9\textit{f}2\textit{g} \\
Ba & 56 & 24\textit{s}22\textit{p}15\textit{d}10\textit{f}2\textit{g}\\
La & 57 & 24\textit{s}22\textit{p}14\textit{d}10\textit{f}2\textit{g}\\
Yb & 70 & 26\textit{s}21\textit{p}14\textit{d}10\textit{f}2\textit{g}\\
Hf & 72 & 25\textit{s}22\textit{p}16\textit{d}10\textit{f}2\textit{g}\\
Hg & 80 & 25\textit{s}21\textit{p}15\textit{d}10\textit{f}2\textit{g}\\
Pb & 82 & 25\textit{s}22\textit{p}16\textit{d}10\textit{f}2\textit{g}\\
Ra & 88 & 26\textit{s}23\textit{p}16\textit{d}11\textit{f}2\textit{g}\\
Ac & 89 & 26\textit{s}24\textit{p}16\textit{d}11\textit{f}2\textit{g}\\
\hline\hline
\end{tabular}
\end{table}

In our previous work \cite{BorIliDzu12} we have examined and compared various schemes for adding electron correlation to the 
Dirac--Hartree--Fock $W_A$ values, and core-polarization contributions to the DFT results. Here, we correct the calculated DHF and 
DFT $W_A$ for core polarization using a scaling parameter, $K_{\rm CP}$. This parameter is obtained from atomic calculations as described 
in the following. 

The main contribution to the matrix elements of the NSD interaction for the 
valence molecular electrons comes from short distances around the heavy nucleus, where the total molecular potential is spherically symmetric to very high precision, and the core of the heavy atom is practically unaffected by the presence of the second atom, justifying our use of the atomic model. The molecular orbitals of the valence electron can thus be expanded in this region, using spherical harmonics centered at
the heavy nucleus,
\begin{equation}
  |\psi_v \rangle= a |s_{1/2} \rangle +b |p_{1/2} \rangle + c|p_{3/2} \rangle + d|d_{3/2} \rangle \dots
\label{eq:psi_v}
\end{equation}
Only $s_{1/2}$ and $p_{1/2}$ terms of this expansion give significant
contribution to the matrix elements of the weak interaction.
These functions can be considered as states of an atomic
 valence electron and are calculated
using standard atomic techniques in two different approximations: one
that includes electron correlation and another that does not.

The single electron DHF Hamiltonian is given by
\begin{equation}
  \hat H_0 = c \bm{\alpha} \cdot \mathbf{p} + (\bm{\beta} -1)c^2 -
  \frac{Z}{r} + V_e(r), 
\label{eq:h0}
\end{equation}
where $\bm{\alpha}$ and $\bm{\beta}$ are the Dirac matrices and $V_e(r)$ is the self-consistent DHF potential due to atomic electrons.

The self-consistent DHF procedure is first performed for the closed shell ion, from
which the valence electron is removed. Then the core potential $V_{\rm
  DHF}^{N-N_v}$ is frozen and the valence $s_{1/2}$ and $p_{1/2}$ states
are calculated by solving the DHF equation for the valence electron,
\begin{equation}
(\hat H_0 - \epsilon_v) \psi_v=0,
\label{eq:DHF}
\end{equation}
where $\hat H_0$ is given by (\ref{eq:h0}).

The core polarization can be understood as the change of the
self-consistent DHF potential due to the effect of the extra term (the
weak interaction operator $\hat H_{\rm A}$) in
the Hamiltonian. The inclusion of the core polarization in
a self-consistent way is equivalent to the random-phase
approximation (RPA, see, e.g.~\cite{DzuFlaSil87}). The change in the DHF
potential is found by solving the RPA-type equations self-consistently
for all states in the atomic core, 
\begin{equation}
(\hat H_0 - \epsilon_c) \delta \psi_c = (-\hat H_{\rm A} + \delta V_{\rm A})\psi_c.
\label{eq:RPA}
\end{equation}
Here, $\hat H_0$ is the DHF Hamiltonian (\ref{eq:h0}), index $c$
enumerates the states in the core, $\delta \psi_c$ is the correction to the
core state $c$ due to weak interaction $\hat H_{\rm A}$, and $\delta V_{\rm A}$
is the correction to the self-consistent core potential due to the
change of all core functions. Once $\delta V_{\rm A}$ is found, the core polarization can be
included into a matrix element for valence states $v$ and $w$ via the
redefinition of the weak interaction Hamiltonian,
\begin{equation}
\langle v |\hat H_{\rm A}| w \rangle \rightarrow \langle v |\hat H_{\rm
  A} + \delta V_{\rm A}| w \rangle. 
\label{eq:meRPA}
\end{equation}
We then obtain the scaling parameter for core-polarization effects, $K_{\rm CP}$, from

\begin{equation}
  K_{\rm CP} = \frac{\langle \psi^{\rm DHF}_{ns_{1/2}}  | \hat H_{\rm A} + \delta
    V_{\rm A}
    |\psi^{\rm DHF}_{n^{\prime}p_{1/2}} \rangle}{\langle \psi^{\rm DHF}_{ns_{1/2}}|
    \hat H_A |\psi^{\rm DHF}_{n^{\prime}p_{1/2}} \rangle} .
\label{eq:K_CP}
\end{equation}

It should be noted that for the group 14 fluorides we have only calculated the correlations between the valence electrons and the core; the correlations between the valence $ns$ and $np$ electrons are not included.

As the final recommended value for the $W_A$ parameter we take an average of $W_A$(DHF)$K_{\rm CP}$ and $W_A$(DFT)$K_{\rm CP}$. The estimate of the accuracy of the final results in our previous work \cite{BorIliDzu12} has shown that it is about 15\% for the molecules in the $^2\Sigma_{1/2}$ electronic state and 20-30\% for the $^2\Pi_{1/2}$ state.

\section{Results and Discussion}

Table II contains the $W_A$ constants obtained for neutral molecules, where other theoretical results are available. To the best of our knowledge this table sums up all the existing calculations of the $W_A$ parameter for neutral diatomic molecules. These systems were discussed in our earlier publication \cite{BorIliDzu12}; however our improved results presented here are are lower than the values shown in Ref. \cite{BorIliDzu12} due to the more accurate treatment of the finite nuclear size and a better treatment of the basis set. This decrease is only $5\%$ for the lighter systems but up to $20\%$ for the heaviest ones. The contribution of correlation on the DFT level to the calculated $W_A$ constants is very small, and for the majority of the molecules considered here it is less than 10\%.

\begin{table*}
  \caption{Internuclear distances $R_{e}$ (\AA ) (experimental, Ref. \cite{HubHerNIST}, unless referenced otherwise), core-polarization scaling parameters $K_{CP}$, the $P$-odd interaction constants $W_{A}$ (Hz) obtained using DHF and DFT methods, and the final recommended values, $W_A$(Final), compared to earlier results.}
  \label{tab:II}
  \setlength{\tabcolsep}{0.3cm}
  \centering
  \resizebox{\textwidth}{!}{
  \begin{tabular}
    [c]{ccccrrrrrlll}\hline\hline
    & Z & $R_{e}$ (\AA ) &$K_{CP}$& \multicolumn{3}{c}{$W_{A}$ (Hz)} &\qquad& \multicolumn{3}{c}{Previous results}\\\cline{5-7}\cline{9-11}%

    \multicolumn{1}{l}{} & \multicolumn{1}{l}{} &
    \multicolumn{1}{l}{}& \multicolumn{1}{l}{} &DHF&DFT&Final&& $W_{A}$ (Hz) & Method & Ref.\\\hline
    \multicolumn{1}{l}{MgF} & \multicolumn{1}{l}{12} & \multicolumn{1}{l}{1.750} &1.2& 3.84&4.36&4.92 && 3.9 &
    ZORA\footnote{Quasirelativistic two-component zero-order regular approximation}+HF & \cite{IsaBer12}\\
    \multicolumn{1}{l}{} & \multicolumn{1}{l}{} &
    \multicolumn{1}{l}{} &\multicolumn{1}{l}{} &   & & && 4.9 & ZORA+DFT(B3LYP\footnote{B3LYP functional}) & \cite{IsaBer12}\\
    
    \multicolumn{1}{l}{CaF} & \multicolumn{1}{l}{20} & \multicolumn{1}{l}{1.967} &1.3& 8.03&8.23&10.6 && 8.0 &
    ZORA+HF & \cite{IsaBer12}\\
    \multicolumn{1}{l}{} & \multicolumn{1}{l}{} &
    \multicolumn{1}{l}{} &\multicolumn{1}{l}{} &   & & && 9.2 & ZORA+DFT(B3LYP) & \cite{IsaBer12}\\
    
    \multicolumn{1}{l}{MgBr} & \multicolumn{1}{l}{35} & \multicolumn{1}{l}{2.360}&1.2&8.24& 17.4&16.2&&18 &Semiempirical & \cite{DemCahMur08}\\
    
    \multicolumn{1}{l}{SrF} & \multicolumn{1}{l}{38} & \multicolumn{1}{l}{2.075}&1.3&40.7 & 39.0 &51.8& & 65 &Semiempirical & \cite{DemCahMur08}\\
    \multicolumn{1}{l}{} & \multicolumn{1}{l}{} &
    \multicolumn{1}{l}{} &\multicolumn{1}{l}{} &   & & && 39 & ZORA+HF & \cite{IsaBer12}\\
    \multicolumn{1}{l}{} & \multicolumn{1}{l}{} &
    \multicolumn{1}{l}{} &\multicolumn{1}{l}{} &   & & && 46 & ZORA+DFT(B3LYP) & \cite{IsaBer12}\\

    \multicolumn{1}{l}{ZrN} & \multicolumn{1}{l}{40} & \multicolumn{1}{l}{1.696}&1.2 &60.0 & 54.0 &68.5&& 99 &
    Semiempirical & \cite{DemCahMur08}\\

    \multicolumn{1}{l}{BaF} & \multicolumn{1}{l}{56} & \multicolumn{1}{l}{2.162} &1.3& 112.9&111.6&146.0 && 164 &
    Semiempirical & \cite{DemCahMur08}\\
    \multicolumn{1}{l}{} & \multicolumn{1}{l}{} &
    \multicolumn{1}{l}{} &\multicolumn{1}{l}{} &   & & && 135 & DHF & \cite{NayDas09}\\
    \multicolumn{1}{l}{} & \multicolumn{1}{l}{}  &
    \multicolumn{1}{l}{} & \multicolumn{1}{l}{} &  & & && 160 & 4c-RASCI\footnote{Fully relativistic restricted active space configuration
      interaction method.} & \cite{NayDas09}\\
      
      \multicolumn{1}{l}{} & \multicolumn{1}{l}{}  &
    \multicolumn{1}{l}{} & \multicolumn{1}{l}{} &  & & && 111 & RECP+

SCF\footnote{Relativistic effective core potential (RECP) combined with SCF} &
    \cite{KozTitMos97}\\
    \multicolumn{1}{l}{} & \multicolumn{1}{l}{}  &
    \multicolumn{1}{l}{} & \multicolumn{1}{l}{} &  & & && 181 & RECP+

SCF+EO\footnote{RECP combined with SCF 
      and an effective operator to account for core-valence correlations.} &
    \cite{KozTitMos97}\\
    \multicolumn{1}{l}{}  & \multicolumn{1}{l}{} &
    \multicolumn{1}{l}{}  & \multicolumn{1}{l}{} && &  && 175 & RECP+RASSCF+EO\footnote{
      RECP combined with restricted active 
      space SCF approach and an effective operator to 
      account for core-valence correlations.} &
    \cite{KozTitMos97}\\
    \multicolumn{1}{l}{} & \multicolumn{1}{l}{} &
    \multicolumn{1}{l}{}  &\multicolumn{1}{l}{} & &  &  && 210-240 & Semiempirical & \cite{KozLab95}\\
    \multicolumn{1}{l}{} & \multicolumn{1}{l}{} &
    \multicolumn{1}{l}{} &\multicolumn{1}{l}{} &   & & && 111& ZORA+HF & \cite{IsaBer12}\\
    \multicolumn{1}{l}{} & \multicolumn{1}{l}{} &
    \multicolumn{1}{l}{} &\multicolumn{1}{l}{} &   & & && 119 & ZORA+DFT(B3LYP) & \cite{IsaBer12}\\
    \multicolumn{1}{l}{} & \multicolumn{1}{l}{} &
    \multicolumn{1}{l}{} &\multicolumn{1}{l}{} &   & & && 190 & Scaled ZORA+HF\footnote{Scaled by a semiemprirical parameter to estimate spin-polarization contribution} & \cite{IsaBer12}\\

    \multicolumn{1}{l}{LaO} &\multicolumn{1}{l}{57} & \multicolumn{1}{l}{1.825}&1.2&149.4 &146.0 &180.2 && 222 &
    Semiempirical & \cite{DemCahMur08}\\

    \multicolumn{1}{l}{YbF} & \multicolumn{1}{l}{70} & \multicolumn{1}{l}{2.016}&1.2 &466.5 & 494.0&576.3 && 729 &
    Semiempirical & \cite{DemCahMur08}\\
    \multicolumn{1}{l}{} & \multicolumn{1}{l}{}  &
    \multicolumn{1}{l}{} &\multicolumn{1}{l}{} &&& && 484 & RECP+SCF & \cite{TitMosEzh96}\\
    \multicolumn{1}{l}{}  & \multicolumn{1}{l}{} &
    \multicolumn{1}{l}{} &\multicolumn{1}{l}{} &&&&& 486 & RECP+RASSCF & \cite{TitMosEzh96}\\
    \multicolumn{1}{l}{} & \multicolumn{1}{l}{} &
    \multicolumn{1}{l}{} &\multicolumn{1}{l}{} &&&&& 634 & RECP+RASSCF+EO & \cite{MosKozTit98}\\
    \multicolumn{1}{l}{} & \multicolumn{1}{l}{} &
    \multicolumn{1}{l}{} &\multicolumn{1}{l}{} &   & & && 465 & ZORA+HF & \cite{IsaBer12}\\
    \multicolumn{1}{l}{} & \multicolumn{1}{l}{} &
    \multicolumn{1}{l}{} &\multicolumn{1}{l}{} &   & & && 610 & Scaled ZORA+HF & \cite{IsaBer12}\\

    \multicolumn{1}{l}{HgF} & \multicolumn{1}{l}{80} & \multicolumn{1}{l}{2.025\footnote{CCSD(T), present calculations}}& 1.1 &3024.7& 2724.5& 3162.1 && 2700 & Semiempirical &
    \cite{KozLab95}\\

    \multicolumn{1}{l}{PbF} & \multicolumn{1}{l}{82} & \multicolumn{1}{l}{2.078
      }&1.1 &$-$1139.9& $-$1167.6 &$-$1269.1 && $-720$ &
    Semiempirical & \cite{DmiKhaKoz92}\\
    \multicolumn{1}{l}{} & \multicolumn{1}{l}{}  &
    \multicolumn{1}{l}{}  & \multicolumn{1}{l}{} && & && $-950\pm$300 & Semiempirical &
    \cite{KozFomDmi87}\\

    \multicolumn{1}{l}{}  & \multicolumn{1}{l}{} &
    \multicolumn{1}{l}{}  &\multicolumn{1}{l}{} & & & && $-990$ & RECP+SODCI\footnote{
      RECP combined with spin-orbit direct configuration
interaction} &
    \cite{BakPetTit10}\\

    \multicolumn{1}{l}{RaF} & \multicolumn{1}{l}{88} & \multicolumn{1}{l}{2.255\footnotemark[6]}&1.2 &1363.9  & 1371.4 & 1641.2& & 1300 &
    ZORA+HF & \cite{IsaHoeBer10}\\
    \multicolumn{1}{l}{} & \multicolumn{1}{l}{} &
    \multicolumn{1}{l}{} &\multicolumn{1}{l}{} &   & & && 1420 & ZORA+DFT(B3LYP) & \cite{IsaBer12}\\
    \multicolumn{1}{l}{} & \multicolumn{1}{l}{} &
    \multicolumn{1}{l}{} &\multicolumn{1}{l}{} &   & & && 2100 & Scaled ZORA+HF & \cite{IsaBer12}\\\hline\hline
  \end{tabular}
}
\end{table*}

For MgBr, SrF, ZrN, BaF, LaO, YbF, HgF, and PbF a number of semiempirical calculations were performed \cite{KozFomDmi87,DmiKhaKoz92,KozLab95,DemCahMur08}. For the lighter systems our $W_A$ constants are about 10-20$\%$ lower than the semiempirical values, for HgF and PbF our values are higher. This discrepancy requires still
some future attention and will be resolved only if a more a accurate electron correlation treatment becomes available. There is already progress in this
direction \cite{timofleig}.

Our uncorrelated DHF  values are in very good agreement with those from Refs. \cite{TitMosEzh96,KozTitMos97,NayDas09,IsaBer12}. In particular, the recently published quasirelativistic two-component zero-order regular approximation (ZORA) HF $W_A$ constants for group 2 fluorides \cite{IsaBer12} do not differ by more than a few percent from the values obtained here. This shows that the performance of the ZORA approximation compared to the 4-component Dirac Hamiltonian is indeed very good for the properties studied here. Ref. \cite{IsaBer12} is also the only other publication where the DFT approach was used. The slightly larger (compared to the difference between the HF results) discrepancies between our DFT values and those from Ref. \cite{IsaBer12} can be attributed to a different choice of functional. 

Our final results, corrected for the core polarization contribution, can be compared to the values obtained in Refs. \cite{KozTitMos97,MosKozTit98,IsaBer12}. In Refs. \cite{KozTitMos97,MosKozTit98}, the authors correct for the core polarization contribution by introducing an effective operator (EO), formed in the framework of the atomic many-body perturbation theory. This method was applied to the BaF and the YbF molecules, increasing the calculated $W_A$ constant by 70\% and by 30\%, respectively, compared to the SCF values. The second attempt to correct for the core polarization contribution is from Ref. \cite{IsaBer12}, where a scaling parameter derived from a semiempirical molecular model of Kozlov \cite{Koz85_2} was applied to the HF $W_A$ constants of BaF, YbF, and RaF, also increasing their magnitude. The results obtained using the two approaches are in very good agreement with each other. Our final values are lower by about 20\% compared to those of Refs. \cite{KozTitMos97,MosKozTit98,IsaBer12}, as the magnitude of the scaling parameters used in this work is 1.1-1.3. Overall, our results are in good agreement with the majority of the previous investigations.

\begin{table}
 \caption{Internuclear distances $R_{e}$ (\AA ) (experimental, Ref. \cite{HubHerNIST}, unless referenced otherwise), core-polarization scaling parameters $K_{CP}$, the $P$-odd interaction constants $W_{A}$ (Hz) obtained using DHF and DFT, and the final recommended values, $W_A$(Final) for neutral diatomic molecules.}
  \centering
  \begin{tabular}{llllrrr}
    \hline\hline
     & $Z$ &  $R_e$ (\AA )& $K_{CP}$ & \multicolumn{3}{c}{$W_A$ (Hz)}  \\
    \cline{5-7}
                   &                        &         &     & DHF & DFT & Final                     \\
    \hline
    \\
    \multicolumn{7}{c}{\textit{Group 3 oxides $(^{2}\Sigma_{1/2})$}} \\
    ScO  &  21 &     1.668  &1.2 & 11.5  & 10.2 &13.0\\
    YO  &39&1.790  & 1.2 &54.9 & 53.6    & 65.1  \\
    LaO &  57 &    1.825  &1.2&150.6& 147.8& 179.0 \\
    AcO   &  89   &1.962\footnote{CCSD(T), present calculations}     &1.2& 2007.9 &1973.5& 2388.9    \\
    \\    
    \multicolumn{7}{c}{\textit{Group 4 nitrides $(^{2}\Sigma_{1/2})$}} \\  
    TiN & 22&     1.582     & 1.2 & 11.6& 10.2&13.1 \\ 
    ZrN  &  40    &   1.696  &1.2 &  59.9& 53.9  & 68.3  \\
    HfN & 72    &   1.736\footnotemark[1] &1.2 & 772.2& 715.1  & 892.4    \\
\\
    \multicolumn{7}{c}{\textit{Group 12 fluorides $(^{2}\Sigma_{1/2})$}} \\
    ZnF   &  30   & 1.766\footnote{Ref. \cite{FloMcLZiu06}}     &1.1 &52.1  &57.4 & 60.3  \\
    CdF &  48    & 1.991  & 1.1 & 220.3  &  227.4   &  246.3    \\
    HgF &  80  &2.025 &1.1&3024.7 & 2724.5 &3162.1   \\
    \\
    \multicolumn{7}{c}{\textit{Group 14 fluorides $(^{2}\Pi_{1/2})$}} \\  
    SiF   &  14    &1.601     &1.1  & $-$0.04 &$-$0.05    & $-$0.05  \\
    GeF&32&    1.745       &1.1  &$-$2.74  &$-$3.17  &$-$3.25 \\ 
    SnF &  50   &1.944     &1.1  & $-$30.0 &$-$34.1 &$-$35.3    \\
    PbF& 82 &  2.078    & 1.1 & $-$1139.9 & $-$1167.6  &$-$1269.1 \\
    
    \hline\hline
  \end{tabular}
  \label{tab:III}
\end{table}
Table III contains the recommended $W_A$ parameters for a number of molecules in the  $^{2}\Sigma_{1/2}$ ground state (group 3 oxides, group 4 nitrides, and group 12 fluorides) and in the $^{2}\Pi_{1/2}$ state (group 14 fluorides). A large fraction of these systems has not been yet investigated in the context of NSD parity violation, and might be suitable for future experimental measurements. 

\begin{figure}
  \centering
  \includegraphics[width=0.5\textwidth]{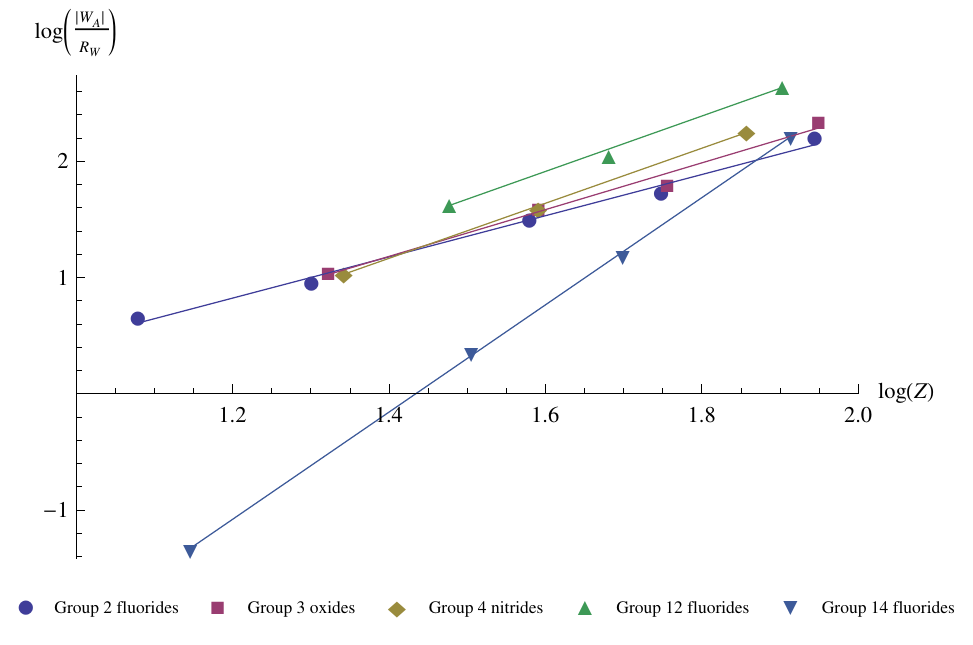}
  \caption{(color online) Scaling of $\log\left(
\frac{W_{A}}{R_{W}}\right)  $ with $\log(Z)$ for the investigated molecules.}
  \label{fig:I}
\end{figure}

The magnitude of $W_{A}$ in the $^2\Sigma_{1/2}$ electronic state is expected to scale as $Z^{2}R_W$ \cite{FlaKri85}, where $R_W$ is the relativistic 
enhancement parameter ($R_W \ge 1$),
\begin{align}
R_W  & =\frac{2\gamma+1}{3}\left(  \frac{a_{B}}{2Zr_{0}A^{1/3}}\right)
^{2-2\gamma}\frac{4}{\left[  \Gamma(2\gamma+1)\right]  ^{2}},\label{Rw}\\
\gamma & =[1-(Z\alpha)^{2}]^{1/2}.\nonumber
\end{align}

In Eq.~(\ref{Rw}), $a_{B}$ is the Bohr radius, $r_{0}=1.2\times10^{-15}$
m, and $\alpha$ is the
fine-structure constant. In Fig. \ref{fig:I} we plot $\log\left(
\frac{|W_{A}|}{R_{W}}\right)  $ as a function of $\log(Z)$ for the five groups of
dimers studied here. A linear fit for each of the groups provides us with the $Z$-exponent $a$ and prefactor $b$ for $|W_{A}| = bR_{W}Z^a$, listed in Table IV for the DHF, DFT, and the final results. The $a$ and $b$ parameters are in good agreement for the DHF and the DFT approaches. For group 2 fluorides and group 3 oxides the scaling is, indeed, close to $Z^{2}$, as suggested by Eq. (10). For group 4 nitrides and  group 12 fluorides the $Z$-dependence is much more advantageous, of $Z^{2.4}$, due to the filling of the lower lying $d$-shell, which expands relativistically and thus increases the effective nuclear charge, leading to an enhancement of relativistic and 
PV effects \cite{AutSieSet02}, and an increase of $W_{A}$. Moreover, even though the group 3 oxides and group 4 nitrides are isoelectronic, nitrogen is more electropositive
than oxygen, thus increasing the electron density at the metal atom. This shows the subtle interplay of electronic effects in these molecules which should be further studied.

In the case of group 14 fluorides, the ground state is $^{2}\Pi_{1/2}$, for which the $W_A$ parameter vanishes in the non-relativistic limit, since in this limit it does not contain the $s$-wave electronic orbital and can not provide the matrix element $\langle s_{1/2}|\bm{\alpha}\rho(\mathbf{r})|p_{1/2}\rangle$. The effect appears due to the mixing of  the $^2\Sigma_{1/2}$ and $^2\Pi_{1/2}$ electronic states by the spin-orbit interaction, which obviously increases with increasing nuclear charge, and gives an extra factor of $Z^2 \alpha^2$ in the $Z$-dependence of $W_A$, as seen in Table IV. However, in this case the prefactor $b$ is four orders of magnitude smaller. Nevertheless, it shows that such molecules should not be discarded for the successful measurement of the anapole moment.

\begin{table}
 \caption{The $Z$-exponent $a$ and the prefactor $b$, derived from $\log\left(
\frac{|W_{A}|}{R_{W}}\right)$ as a function of $\log(Z)$. }
  \centering
  \begin{tabular}{lccccccc}
    \hline\hline
     &  \multicolumn{3}{c}{$a$} && \multicolumn{3}{c}{$\log(b)$}  \\
     \cline{2-4} \cline{6-8}\\
                   &     DHF    &   DFT      &  Final  & & DHF & DFT & Final \\
    \hline                                  \\
    
    Group 2 fluorides &  1.80 &    1.74  &1.77 & &$-$1.45 & $-$1.35 &$-$1.30\\
    Group 3 oxides &1.98&2.05  & 2.01& &$-$1.65 &  $-$1.78 & $-$1.63  \\
    Group 4 nitrides&  2.35 &2.39 &2.37&&$-$2.18& $-$2.30& $-$2.16 \\
    Group 12 fluorides &  2.49   & 2.28 &2.38&&$-$2.12 &$-$1.77& $-$1.90   \\
    Group 14 fluorides &  4.67   &4.57 &4.61&&$-$6.78 &$-$6.56& $-$6.61   \\
    
    \hline\hline
  \end{tabular}
  \label{tab:III}
\end{table}

\section{Conclusions}

We have obtained the much improved $W_A$ parameters for diatomic molecules from relativistic HF and DFT calculations using corrections for core polarization effects. The scaling of the $W_A$ parameter with the nuclear charge $Z$ was examined in different groups and found to be advantageous in group 4 nitrides and group 12 fluorides.
Some of the heavier molecules, such as HgF and PbF, should be excellent candidates for future NSD PV measurements.
The accurate inclusion of electron correlation within a fully relativistic framework in the calculations of the $W_A$ parameters, which still remains a challenge, will be the focus of our future work.

\textbf{Acknowledgements:}
This work was supported by the Marsden Fund (Royal Society of New Zealand), the Australian Research Council, and
the Alexander von Humboldt Foundation (Bonn). M.I. is grateful for the financial support from the Slovak Research and Development Agency (Grant No. APVV-0059-10) and from the Agency of the Ministry of Education, Science, Research and Sport of the Slovak Republic for the Structural Funds of EU (ITMS 26110230082). 
The authors are very grateful to R. Berger and T. Isaev from the TU Darmstadt for their help and fruitful discussions.

%

\end{document}